\documentclass[12pt]{article}
\usepackage{axodraw}

\newcommand{\beq}{\begin{equation}}
\newcommand{\eeq}{\end{equation}}
\newcommand{\beqa}{\begin{eqnarray}}
\newcommand{\eeqa}{\end{eqnarray}}
\newcommand{\beqar}{\begin{eqnarray*}}
\newcommand{\eeqar}{\end{eqnarray*}}

\def\non          {\nonumber}

\def\Tr           {\mbox{\rm Tr}\,}
\def\STr          {\mbox{\rm STr}\,}

\def\ran          {\rangle}
\def\lan          {\langle}

\newcommand{\ga}{\gamma}
\newcommand{\Ga}{\Gamma}

\newcommand{\inn}{\!\cdot\!}

\newcommand{\lam}{\lambda}

\newcommand{\z}{\zeta}

\newcommand{\eg}{{\it e.g.,}\ }
\newcommand{\ie}{{\it i.e.,}\ }
\newcommand{\labell}[1]{\label{#1}} 
\newcommand{\reef}[1]{(\ref{#1})}
\newcommand\prt{\partial}

\newcommand\btheta{\bar{\theta}}

\newcommand\bu{\bar{u}}

\newcommand\bPsi{{\bar \Psi }}

\begin{document}
\baselineskip 18pt%
\begin{titlepage}
\vspace*{1mm}%
\hfill%
\vspace*{15mm}%

\centerline{{\Large {\bf Tachyon Couplings to  Fermion  }}}
\vspace*{5mm}
\begin{center}
{Mohammad R. Garousi}\\
\vspace*{0.2cm}
{ Department of Physics, Ferdowsi University of Mashhad, \\
P.O. Box 1436, Mashhad, Iran}\\
\vspace*{0.1cm}
{ School of Physics, 
 Institute for research in fundamental sciences (IPM), \\
 P.O. Box 19395-5531, Tehran, Iran. 
}\\\vspace*{1.5cm}
\end{center}

\begin{center}{\bf Abstract}\end{center}
\begin{quote}
By fixing the internal CP factor of tachyon and massless Ramond vertex operators in different pictures, we have shown that the internal CP factor of the disk level S-matrix elements of two fermions  and  odd number of tachyon vertex operators in the world volume of non-BPS D-branes/D-brane-anti-D-brane is zero.  We have  calculated  the S-matrix element of two fermions and two tachyons  which has non vanishing  internal CP factor, and found the momentum expansion of  this amplitude.  In the abelian case, we have compared  the   two-fermion-two-tachyon coupling  at low energy  with  the corresponding coupling in the gauge-fixed supersymmetric  tachyon DBI action. The couplings in the two cases are  exactly the same.

\end{quote}
\end{titlepage}
\section{Introduction}
 Study of unstable objects   might shed new light
in understanding properties of string theory in time-dependent
backgrounds \cite{Gutperle:2002ai,Sen:2002in,Sen:2002an,Sen:2002vv,Lambert:2003zr,Sen:2004nf}. It has been
shown by A. Sen that the tachyon DBI action 
\cite{Sen:1999md,Garousi:2000tr,Bergshoeff:2000dq,Kluson:2000iy}  can capture many properties
of the decay of the non-BPS D-branes 
\cite{Sen:2002in,Sen:2002an} around the stable point of the tachyon potential. 

This  action has been proposed in \cite{Garousi:2000tr} to reproduce the leading order terms of the momentum expansion of the disk level tachyon S-matrix elements.  The S-matrix method can be  used to find the tachyon action around the unstable point of non-BPS D-branes/D-brane-anti-D-brane where the higher derivatives of the tachyon are important. However, one may use the resulting action around the stable point where the higher derivative terms are not important. A subtlety  in the tachyon DBI action around unstable point that the S-matrix method dictates  is that   while the massless fields carry identity internal CP matrix,   the tachyon   must carry  $\sigma_1$ and $\sigma_2$  matrices  \cite{Garousi:2008tn}. This subtlety however  may not appear in the tachyon action around the stable point. The WZ part of the effective action of non-BPS D-branes/D-brane-anti-D-brane can also be reproduced exactly by the leading order terms of the corresponding S-matrix elements \cite{Kennedy:1999nn,Billo:1999tv,Garousi:2007fk,Garousi:2008tn}. These couplings has been also found in \cite{Kraus:2000nj,Takayanagi:2000rz} from boundary string field theory.

The world volume fermions has been added into the tachyon DBI action by making the action  supersymmetric \cite{Bergshoeff:2000dq}. If one removes the tachyon field, the action is then the  supersymmetrized DBI  action \cite{Aganagic:1996nn,Aganagic:1997zk}. The 16-component fermions $\theta_1,\,\theta_2$ in this action are related to 32-component fermion $\theta$ as 
\beqa
\theta_1=\frac{1}{2}(1+\Gamma^{11})\theta,\qquad \theta_2=\frac{1}{2}(1-\Gamma^{11})\theta
\eeqa
In static gauge, one choses $\theta_2=0$ \cite{Aganagic:1996nn}. Returning  the tachyon to gauge-fixed action, one finds
\beqa
L&\!\!\!=\!\!\!&-T_pV(T)\sqrt{-\det(\eta_{ab}+F_{ab}-2\btheta_1\ga_a\prt_a\theta_1+\btheta_1\ga^{\mu}\prt_a\theta_1\btheta_1\ga_{\mu}\prt_b\theta_1 + \prt_aT\prt_bT )}\labell{super}
\eeqa
In this paper we would like to find the couplings of tachyon to fermion with the S-matrix method and compare the result with the above action. 

 An outline of the   paper is as follows. In the next section, by explicit calculation of some examples,   we will show that a S-matrix element is independent of the choice of the picture of   the vertex operators when one includes the internal CP factors. This allows one to calculate a S-matrix element in one particular picture and then rewrite the result in another picture in which the vertex operators have the same internal CP matrices as  the CP matrices of the corresponding fields in the effective action, \eg the massless fields in the field theory carry identity CP matrix. In section 2.1, we will also clarify the presence of internal CP matrices $\sigma_1$ and $\sigma_2$ for tachyon  in the tachyon DBI action. In section 3, by specifying the internal matrix of the massless Ramond vertex operator, we will show that the internal CP factor of the S-matrix element of two fermions and odd number of tachyons is zero. In section 4, we will calculate the S-matrix element of two fermions and two tachyons and find a momentum expansion for the amplitude. At one momentum level, we find a coupling which is zero for abelian case. At three momentum level and for abelian case, we find that the coupling is exactly the same as the corresponding coupling in the gauge-fixed supersymmetric tachyon DBI action \reef{super}.

\section{S-matrix elements in different pictures}
The internal CP matrix of an  open string of non-BPS D-brane can be read from the external CP matrix of D-brane-anti-D-brane \cite{Sen:1999mg}. The non-BPS D$_p$-branes of type IIA(B) string theory are defined by projecting D$_p$-brane-anti-D$_p$-brane of type IIB(A) with $(-1)^{F_L}$ where $F_L$ denotes the contribution to the space-time fermion number from the left-moving sector of the string world-sheet \cite{Sen:1999mg}. The open strings of the brane-anti-brane can be labeled by the ${\it external}$  $2\times 2$ Chan-Paton factors:
\beqa
(a):\,\pmatrix{0&0\cr 
0&1},\,\,\,
(b):\,\pmatrix{1&0\cr 
0&0},\,\,\, (c):\,\pmatrix{0&0\cr 
1&0},\,\,\, (d):\,\pmatrix{0&1\cr 
0&0}\labell{M121} \eeqa   
The massless states carry CP factor (a), (b), and the tachyons carry (c) and (d) factors. The projection operator $(-1)^{F_L}$ has no effect on the world-sheet fields, however, using the fact that it exchanges brane with anti-brane, one observes that its effect on the CP matrix $\Lambda$ is the following: 
\beqa
\Lambda\rightarrow \sigma_1\Lambda (\sigma_1)^{-1}, \nonumber\eeqa
where $\sigma_1$  is the Pauli matrix.  The states with CP matrices $I$ and $\sigma_1$ are survived. The massless fields then carry the ${\it internal}$ CP matrix $I$ and the real tachyon of non-BPS D-brane carries the CP factor $\sigma_1$. 

One needs to define the  vertex operator in different pictures to be able to calculate the S-matrix elements. We assign the above internal CP matrices to the vertex operators in the 0 picture. Using the observation   that the picture changing operator on a non-BPS brane naturally comes with $\sigma_3$ \cite{DeSmet:2000je}, one observes that in -1 picture, the internal CP matrix of massless states is $\sigma_3$ and the CP matrix of tachyon is $\sigma_2$. The open strings of $D\bar{D}$ system carry the same internal CP matrices, as well as the external CP matrices \reef{M121}. 

The closed string Ramond-Ramond vertex operator also carries the internal CP matrix. It has been shown in \cite{Garousi:2008tn} that the RR vertex operator in $D\bar{D}$ system in $(-1/2,-1/2)$ picture carries CP matrix $\sigma_3$. For non-BPS branes, there should be  an extra factor of $\sigma_1$ in  the RR vertex operator \cite{Sen:1999mg}. The closed strings in the NSNS sector in $(0,0)$ picture should carry internal CP matrix $I$. Using the internal CP matrices, one can easily check that the CP factor of tree level S-matrix element of one RR, odd number of tachyons and an arbitrary number of closed string NSNS states is zero for $D\bar{D}$ system as expected. Moreover, the internal CP factor of  tree level S-matrix element of one RR, even number of tachyons and an arbitrary number of closed string NSNS states is zero for non-BPS D-brane.


It is well known that in the world volume of BPS D-branes, the S-matrix elements are independent of the choice of the picture of the vertex operators. By explicit calculation of some S-matrix elements, we are going to show that in the world volume of non-BPS D-branes, the S-matrix elements are independent of the choice of the picture of the vertex operators only when they include  the internal CP factor. As a first  example, consider the S-matrix element of one gauge field and two tachyons. In one particular choice of the pictures, it is given by\footnote{We have set $\alpha'=2$ in the string theory side.}
\beqa
{\cal A}&\sim&\sum_{\rm non-cyclic}\int dx_1dx_2dx_3\Tr\lan V_T^{-1}(x_1)V_T^{-1}(x_2)V_A^{0}(x_3)\ran\labell{amp7}
\eeqa
where the vertex operators are
\beqa
V_T^{-1}&=&e^{-\phi}e^{2ik\cdot X}\lam\otimes\sigma_2,\qquad k^2=1/4\nonumber\\
V_A^{0}&=&\xi_{\mu}(\prt X^{\mu}+2ik\inn\psi\psi^{\mu})e^{2ik\cdot X}\lam\otimes I,\qquad k^2=0,\,\xi\inn k=0\,
\eeqa
where $\lam$ is the external $U(N)$ matrix. The internal CP factor for the above amplitude is $\Tr(\sigma_2\sigma_2)=2$. To perform the correlators, one needs  the propagators for the world-sheet fields, \ie
\begin{eqnarray}
\lan X^{\mu}(z)X^{\nu}(w)\ran & = & -\eta^{\mu\nu}\log(z-w) \ , \non \\
\lan \psi^{\mu}(z)\psi^{\nu}(w) \ran & = & -\eta^{\mu\nu}(z-w)^{-1} \ ,\non \\
\lan\phi(z)\phi(w)\ran & = & -\log(z-w) \ .
\labell{prop}\end{eqnarray}
Performing the correlators and using the on-shell conditions, one finds that the integrand is 
$2ik_1\inn\xi x_{12}^{-1}x_{13}^{-1}x_{23}^{-1}$ for both 123 and 132 orderings. It is $SL(2,R)$ invariant. Removing the volume of the $SL(2,R)$ group which amounts  to multiplying the amplitude by $|x_{12}x_{13}x_{23}|$, one finds that for 123 ordering the amplitude is $2ik_1\inn\xi$ and for 132 ordering it is $-2ik_1\inn\xi$, \ie
\beqa
{\cal A}&\sim&2ik_1\inn\xi\left(\Tr(\lam_1\lam_2\lam_3)-\Tr(\lam_1\lam_3\lam_2)\right)
\eeqa
which is symmetric under interchanging the two tachyons. Another choice for the vertex operators is $\lan V_T^{0}(x_2)V_T^{0}(x_2)V_A^{-2}(x_3)\ran$ where $V_A^{-2}=e^{-2\phi}V_A^{0}$. In this case also the internal CP factor   is $\Tr(\sigma_1\sigma_1)=2$. After performing the correlators, one finds that the integrand is exactly as before, \ie $2ik_1\inn\xi x_{12}^{-1}x_{13}^{-1}x_{23}^{-1}$ for both 123 and 132 orderings. So the final result is the same as above.  For another choice of the pictures, the amplitude is given by 
\beqa
{\cal A}&\sim&\sum_{\rm non-cyclic}\int dx_1dx_2dx_3\Tr\lan V_T^{-1}(x_1)V_T^{0}(x_2)V_A^{-1}(x_3)\ran
\eeqa
where the vertex operators are
\beqa
V_T^{0}&=&2ik\inn\psi e^{2ik\cdot X}\lam\otimes\sigma_1\nonumber\\
V_A^{-1}&=&\xi_{\mu}\psi^{\mu}e^{-\phi}e^{2ik\cdot X}\lam\otimes \sigma_3
\eeqa
Performing the correlators and using the on-shell conditions, one finds that the integrand is 
$2ik_1\inn\xi x_{12}^{-1}x_{13}^{-1}x_{23}^{-1}$ for 123 ordering where $x_{23}^{-1}$ is coming from fermion correlator, and it is $2ik_1\inn\xi x_{12}^{-1}x_{13}^{-1}x_{32}^{-1}$ for 132 ordering. This time $x_{32}^{-1}$ is coming from the fermion correlator. Note that in 132 ordering $x_3<x_2$  and in 123 ordering $x_2<x_3$. Fixing the $SL(2,R)$ symmetry as before, one finds
\beqa
{\cal A}&\sim&2ik_1\inn\xi\left(\Tr(\lam_1\lam_2\lam_3)\Tr(\sigma_2\sigma_1\sigma_3)+\Tr(\lam_1\lam_3\lam_2)\Tr(\sigma_2\sigma_3\sigma_1)\right)
\eeqa
Obviously, without considering the internal CP factors, the two amplitudes are not the same. Using the fact that $\sigma_1\sigma_2=-\sigma_2\sigma_1$, the two amplitudes are identical when considering the CP factors. This S-matrix element is consistent with the coupling $\Tr(D_{\mu}TD^{\mu} T)$.  Using the fact that the effective field theory of non-BPS D-branes should be reduced to the field theory of BPS D-branes when tachyon is set to zero, and the fact that there is no internal CP factor for the field theory of  BPS D-branes,  one realizes that the gauge field in the effective field theory of non-BPS D-branes must carry identity matrix. So in the coupling $\Tr(D_{\mu}TD^{\mu} T)$
the gauge field carries identity  matrix and tachyons carry internal matrix $\sigma_2$ or $\sigma_1$.

\subsection{S-matrix element of four tachyons} 

The   S-matrix element of four tachyons in which two of them are in 0 picture and the other two are in -1 picture are given by one of the following amplitudes:
\beqa
{\cal A}_s&\sim&\sum_{\rm non-cyclic}\int dx_1dx_2dx_3dx_3\Tr\lan V_T^{-1}(x_1)V_T^{-1}(x_2)V_T^{0}(x_3)V_T^{0}(x_4)\ran\nonumber\\
{\cal A}_t&\sim&\sum_{\rm non-cyclic}\int dx_1dx_2dx_3dx_3\Tr\lan V_T^{-1}(x_1)V_T^{0}(x_2)V_T^{0}(x_3)V_T^{-1}(x_4)\ran\nonumber\\
{\cal A}_u&\sim&\sum_{\rm non-cyclic}\int dx_1dx_2dx_3dx_3\Tr\lan V_T^{-1}(x_1)V_T^{0}(x_2)V_T^{-1}(x_3)V_T^{0}(x_4)\ran\nonumber
\eeqa
where  the indexes  $s,t,u$ stand for the Mandelstam  variables which are \beqa
s&=&-(k_1+k_2)^2\,\,,\nonumber\\
t&=&-(k_2+k_3)^2\,\,,\nonumber\\
u&=&-(k_1+k_3)^2\,\,.\labell{mandel} \nonumber\eeqa 
 and   satisfy the constraint \beqa
s+t+u&=&-1\labell{con1}\,\,. \eeqa
Before performing the correlators, let us see how much we know about the amplitude. Each of the above amplitudes has pole in $s$, $t$ and $u$ channels. Using the observation that the gauge field in the effective field theory must have identity internal matrix, one realizes that ${\cal A}_s,\,{\cal A}_t$,${\cal A}_u$ should have massless pole only in $s$-channel, $t$-channel, $u$-channel, respectively. This is consistent with the above constraint which indicates that one can not send all the Mandelstam variables to zero. So $s,t,u$ channels of effective field theory should be corresponding to the following expansions:
 \beqa s-{\rm
channel}:&&\lim_{s\rightarrow 0\,,t,u\rightarrow -1/2}{\cal A}_s\nonumber\\
t-{\rm channel}:&&\lim_{t\rightarrow 0\,,s,u\rightarrow
-1/2}{\cal A}_t\labell{lim1}\\
u-{\rm channel}:&&\lim_{u\rightarrow 0\,,s,t\rightarrow
-1/2}{\cal A}_u\nonumber
\eeqa
From the field theory point of view, we know that  the S-matrix element of four tachyons must have massless pole in all channels. So the correct S-matrix element should be the sum of the three amplitudes, \ie ${\cal A}_s+{\cal A}_t+ {\cal A}_u$.
 
The   S-matrix element of four tachyons can also be given  by the following:
\beqa
{\cal A}&\sim&\sum_{\rm non-cyclic}\int dx_1dx_2dx_3dx_3\Tr\lan V_T^{0}(x_1)V_T^{0}(x_2)V_T^{0}(x_3)V_T^{-2}(x_4)\ran\nonumber
\eeqa
In this case the amplitude  can have massless pole  in all $s,t,u$ channels. However, the constraint \reef{con1} does not allow us to send $s,t,u$ to zero at the same time. So $s,t,u$ channels of effective field theory in this case should be corresponding to the following expansions:
 \beqa s-{\rm
channel}:&&\lim_{s\rightarrow 0\,,t,u\rightarrow -1/2}{\cal A}\nonumber\\
t-{\rm channel}:&&\lim_{t\rightarrow 0\,,s,u\rightarrow
-1/2}{\cal A}\labell{lim2}\\
u-{\rm channel}:&&\lim_{u\rightarrow 0\,,s,t\rightarrow
-1/2}{\cal A}\nonumber
\eeqa  
To have massless pole in all channels, one has to consider the sum of the above expansions, \ie $\lim_{s\rightarrow 0\,,t,u\rightarrow -1/2}{\cal A}+\lim_{t\rightarrow 0\,,s,u\rightarrow
-1/2}{\cal A}+\lim_{u\rightarrow 0\,,s,t\rightarrow
-1/2}{\cal A}$.

Now let us perform the correlators in the  amplitudes. Performing the correlators, one finds that the integrand is $SL(2,R)$ invariant. Removing this symmetry by fixing  $ x_1=0,\, x_3=1,\, x_4=\infty$ for 1234 ordering, $ x_1=0,\, x_4=1,\, x_3=\infty$ for 1243 ordering and so on, one finds
\beqa {\cal A}_s&\sim&k_1\inn k_2\left(A_s\frac{\Ga(-2t)\Ga(-1-2s)}{\Ga(-1-2t-2s)}+
B_s\frac{\Ga(-1-2s)\Ga(-2u)}{\Ga(-1-2s-2u)}+
C_s\frac{\Ga(-2t)\Ga(-2u)}{\Ga(-2t-2u)}\right) \nonumber\eeqa
 The
coefficients $A_s,B_s,C_s$ are 
\beqa A_s&=&\frac{1}{4}\left(\frac{}{}\Tr(\lam_1\lam_2\lam_3\lam_4)\Tr(\sigma_2\sigma_2\sigma_1\sigma_1)+\Tr(\lam_1\lam_4\lam_3\lam_2)\Tr(\sigma_2\sigma_1\sigma_1\sigma_2)\right)\,\,,\nonumber\\
B_s&=&\frac{1}{4}\left(\frac{}{}\Tr(\lam_1\lam_3\lam_4\lam_2)\Tr(\sigma_2\sigma_1\sigma_1\sigma_2)+\Tr(\lam_1\lam_2\lam_4\lam_3)\Tr(\sigma_2\sigma_2\sigma_1\sigma_1)\right)\,\,,\nonumber\\
C_s&=&\frac{1}{4}\left(\frac{}{}\Tr(\lam_1\lam_4\lam_2\lam_3)\Tr(\sigma_2\sigma_1\sigma_2\sigma_1)+\Tr(\lam_1\lam_3\lam_2\lam_4)\Tr(\sigma_2\sigma_1\sigma_2\sigma_1)\right)\,\,.\labell{phase0}\nonumber\eeqa 
Using the identity $4k_1\inn k_2=-1-2s$, one can write the amplitude as
\beqa
{\cal A}
_s&\sim&A_s\frac{\Ga(-2s) \Ga(-2t)}{\Ga(-1-2s-2t)}+
B_s\frac{\Ga(-2s)\Ga(-2u)}{\Ga(-1-2s-2u)}-
C_s\frac{\Ga(-2t)\Ga(-2u)}{\Ga(-1-2t-2u)}\labell{am6}
\eeqa
Performing the  trace over the internal  CP  matrices, one can easily observes that the  amplitude is symmetric  under interchanging the tachyons. However, the expansion $s\rightarrow 0,\,(t,u)\rightarrow -1/2$ is symmetric only under $1\leftrightarrow 2$ and $3\leftrightarrow 4$. 

Similarly, for the  amplitudes ${\cal A}_u$ and ${\cal A}_t$, one finds
\beqa 
{\cal A} _u&\sim&-A_u\frac{\Ga(-2s)
\Ga(-2t)}{\Ga(-1-2s-2t)}+
B_u\frac{\Ga(-2u)\Ga(-2s)}{\Ga(-1-2u-2s)}+
C_u\frac{\Ga(-2u)\Ga(-2t)}{\Ga(-1-2u-2t)}\nonumber\\
{\cal A}_t&\sim&A_t\frac{\Ga(-2t)
\Ga(-2s)}{\Ga(-1-2t-2s)}-
B_t\frac{\Ga(-2s)\Ga(-2u)}{\Ga(-1-2s-2u)}+C_t
\frac{\Ga(-2t)\Ga(-2u)}{\Ga(-1-2t-2u)}
\labell{a717}\eeqa 
where
\beqa A_u&=&\frac{1}{4}\left(\frac{}{}\Tr(\lam_1\lam_2\lam_3\lam_4)\Tr(\sigma_2\sigma_1\sigma_2\sigma_1)+\Tr(\lam_1\lam_4\lam_3\lam_2)\Tr(\sigma_2\sigma_1\sigma_2\sigma_1)\right)\,\,,\nonumber\\
B_u&=&\frac{1}{4}\left(\frac{}{}\Tr(\lam_1\lam_3\lam_4\lam_2)\Tr(\sigma_2\sigma_2\sigma_1\sigma_1)+\Tr(\lam_1\lam_2\lam_4\lam_3)\Tr(\sigma_2\sigma_1\sigma_1\sigma_2)\right)\,\,,\nonumber\\
C_u&=&\frac{1}{4}\left(\frac{}{}\Tr(\lam_1\lam_4\lam_2\lam_3)\Tr(\sigma_2\sigma_1\sigma_1\sigma_2)+\Tr(\lam_1\lam_3\lam_2\lam_4)\Tr(\sigma_2\sigma_2\sigma_1\sigma_1)\right)\,\,.\nonumber\eeqa 
\beqa A_t&=&\frac{1}{4}\left(\frac{}{}\Tr(\lam_1\lam_2\lam_3\lam_4)\Tr(\sigma_2\sigma_1\sigma_1\sigma_2)+\Tr(\lam_1\lam_4\lam_3\lam_2)\Tr(\sigma_2\sigma_2\sigma_1\sigma_1)\right)\,\,,\nonumber\\
B_t&=&\frac{1}{4}\left(\frac{}{}\Tr(\lam_1\lam_3\lam_4\lam_2)\Tr(\sigma_2\sigma_1\sigma_2\sigma_1)+\Tr(\lam_1\lam_2\lam_4\lam_3)\Tr(\sigma_2\sigma_1\sigma_2\sigma_1)\right)\,\,,\nonumber\\
C_t&=&\frac{1}{4}\left(\frac{}{}\Tr(\lam_1\lam_4\lam_2\lam_3)\Tr(\sigma_2\sigma_2\sigma_1\sigma_1)+\Tr(\lam_1\lam_3\lam_2\lam_4)\Tr(\sigma_2\sigma_1\sigma_1\sigma_2)\right)\,\,.\nonumber\eeqa
The above amplitudes, without considering the internal CP factors, have been found in \cite{Garousi:2002wq}. Performing the internal CP factors, one observes that the above amplitudes are   symmetric under interchanging the tachyons. However, the expansion $t\rightarrow 0,\,(s,u)\rightarrow -1/2$ for ${\cal A}_t$ is symmetric only under $2\leftrightarrow 3$ and $1\leftrightarrow 4$. Similarly for ${\cal A}_u$. 
However, the combination ${\cal A}_s+{\cal A}_t+{\cal A}_u$ with the expansion \reef{lim1} is symmetric under interchanging the tachyons.

For the amplitude ${\cal A}$, one finds the integrand for 1234 ordering to be
\beqa
&&|x_{12}|^{4k_1\cdot k_2}|x_{13}|^{4k_1\cdot k_3}|x_{14}|^{4k_1\cdot k_4}|x_{23}|^{4k_2\cdot k_2}|x_{24}|^{4k_2\cdot k_4}|x_{34}|^{4k_3\cdot k_4}\nonumber\\
&&\times\left((4k_1\inn k_2)^2x_{12}^{-1}x_{34}^{-1}-(4k_1\inn k_3)^2x_{13}^{-1}x_{24}^{-1}+(4k_2\inn k_3)^2x_{14}^{-1}x_{23}^{-1}\right) 
\eeqa
where the second line is the world sheet fermion correlator. Fixing $x_1=0, \, x_3=1,\, x_4=\infty$, one finds the integral to be $\Ga(-2t)\Ga(-2s)/\Ga(-1-2t-2s)$. Similarly for the other orderings. The final result is
\beqa
{\cal A}&\sim&\alpha\frac{\Ga(-2t)
\Ga(-2s)}{\Ga(-1-2t-2s)}
+\beta\frac{\Ga(-2s)\Ga(-2u)}{\Ga(-1-2s-2u)}+\gamma
\frac{\Ga(-2t)\Ga(-2u)}{\Ga(-1-2t-2u)}
\eeqa
where
\beqa \alpha&=&\frac{1}{2}\left(\frac{}{}\Tr(\lam_1\lam_2\lam_3\lam_4)+\Tr(\lam_1\lam_4\lam_3\lam_2)\right)\,\,,\nonumber\\
\beta&=&\frac{1}{2}\left(\frac{}{}\Tr(\lam_1\lam_3\lam_4\lam_2)+\Tr(\lam_1\lam_2\lam_4\lam_3)\right)\,\,,\nonumber\\
\gamma&=&\frac{1}{2}\left(\frac{}{}\Tr(\lam_1\lam_4\lam_2\lam_3)+\Tr(\lam_1\lam_3\lam_2\lam_4)\right)\,\,.\labell{abg}\eeqa 
Note that the internal CP factor of   amplitude ${\cal A}$ is $\Tr(\sigma_1\sigma_1\sigma_1\sigma_1)=2$ for any ordering of the vertex operators. The above amplitude is symmetric under interchanging the tachyons.

Using the identities $\sigma_1\sigma_2=-\sigma_2\sigma_2$ and $\Tr(\sigma_1\sigma_1\sigma_2\sigma_2)=2$, one can easily check that ${\cal A}_s={\cal A}_u={\cal A}_t={\cal A}$, as expected. This indicates that the expansion \reef{lim2} is the only possible  expansion for ${\cal A}$, because the only  expansion for, say,  ${\cal A}_s$ is \reef{lim1}.  Comparing the amplitudes ${\cal A}_s+{\cal A}_u+{\cal A}_t$  in which the trace over the internal  CP matrices does not perform, with the S-matrix element of four transverse scalars, one realizes that the expansion \reef{lim1} is very similar to the low energy expansion of the S-matrix element of four scalars \cite{Garousi:2003pv}.  Whereas the expansion \reef{lim2} in which there is no CP factor is not similar to the low energy expansion of the scalar amplitude. This observation has been used in \cite{Garousi:2008tn} to write the four tachyon couplings in the non-abelian tachyon DBI form in which the tachyons carry internal CP matrices, \eg for abelian case, it is
\beqa
L_{DBI}&=&-\frac{T_p}{2}\STr\left(V(T^iT^i)\sqrt{1+\frac{1}{2}[T^i,T^j][T^j,T^i]}\right.\label{DBI}\\
&&\left.\qquad\qquad\qquad\times\sqrt{-\det(\eta_{ab}+2\pi\alpha'F_{ab}+2\pi\alpha'\prt_aT^i(Q^{-1})^{ij}\prt_bT^j)}\frac{}{}\right)\nonumber
\eeqa
where
\beqa
Q^{ij}&=&\delta^{ij}-i[T^i,T^j]\nonumber
\eeqa
 The superscripts  $i,j=1,2$ and there is no sum over them. In above, $T^1=T\sigma_1$ and $T^2=T\sigma_2$. After expanding the square roots, one should choose two of the tachyons to be $T^2$ and the others to be $T^1$, and then performs the symmetric trace which is symmetric between $\prt_aT^i$, $[T^i,T^j]$ and the individual $T^i$ of the tachyon potential. This simplifies the first square root to be $1+[T^1,T^2][T^2,T^1]/4$.  For the first terms, the CP matrix of tachyon in the rest is $\sigma_1$ and/or $\sigma_2$. 
For the second terms, on the other hand,  the internal CP matrix of all other terms is $\sigma_1$ because the two  $\sigma_2$'s   appear   in $[T^1,T^2][T^2,T^1]/4$. For example, for these terms $Q^{ij}=\delta^{ij}$. Around the stable point of the tachyon potential, $T\rightarrow \infty$, one can approximate $1+[T^1,T^2][T^2,T^1]/4\sim [T^1,T^2][T^2,T^1]/4$.
 Hence,  one can  approximate the  action \reef{DBI} with the usual tachyon DBI action in which the tachyon potential is $T^4V(T^2)$. The expansion of $V(TT)$ up to forth order of tachyon is consistent with $e^{-\pi TT/2}$ so  as $T\rightarrow \infty$, the potential $T^4V(T^2)\rightarrow 0$, as expected  in the tachyon condensation of a non-BPS D-brane. 

Having found that the S-matrix elements in different pictures are identical when the internal CP factors are included, we now turn to the calculation of the S-matrix elements involving tachyon and fermion.


\section{The $\bPsi\Psi T^{2n+1}$ amplitude}

 The three point coupling between two  Ramond vertex operators  and one  gauge field vertex operator in the world volume of BPS D-brane  is given by \cite{jp}
 \beqa
{\cal A}^{\bPsi,\Psi,A} & \sim & \bu_1^A\ga^{\mu}_{AB}u_2^B\xi_{\mu}\left(\Tr(\lam_1\lam_2\lam_3)-\Tr(\lam_1\lam_3\lam_2)\right)\labell{amp2}
\eeqa
where $u^A$ is (commuting) 10-dimensional Majorana-Weyl  wave function. In non-BPS D-brane case also  the coupling is  given  by the above amplitude. To fix our notation and specify the internal CP matrix of the Ramond vertex operators, we calculate this amplitude here again. The amplitude is given at the world-sheet level by 
 the following correlation function:
\begin{eqnarray}
{\cal A}^{\bPsi,\Psi,A} & \sim &\sum_{\rm non-cyclic} \int dx_1dx_2dx_3
 \Tr\lan 
V_{\bPsi}^{(-1/2)}(x_1)V_{\Psi}^{(-1/2)}(x_2)V_{A}^{(-1)}(x_3)\ran\labell{cor10}\eeqa
  The internal CP matrix of the Ramond  vertex operators should be defined as
\beqa
V_{\bPsi}^{(-1/2)}&=&\bu^Ae^{-\phi/2}S_A\,e^{2ik.X}\lam\otimes \sigma_3\nonumber\\
V_{\Psi}^{(-1/2)}&=&u^Ae^{-\phi/2}S_A\,e^{2ik.X}\lam\otimes I
\eeqa
to give non-zero result for the above amplitude. The above internal matrices are consistent with the internal CP matrix of the RR vertex operator in $(-1/2,-1/2)$ picture which is $\sigma_3$.   The internal CP factor of the above amplitude is $\Tr(\sigma_3\sigma_3)=1$ for any ordering of the open string vertex operators. The amplitude \reef{amp2} has been found in \cite{jp} by considering 123 and 213 orderings. We would like to consider   123 and 132 orderings here. For 123 ordering one needs the following correlators \cite{jp}:
 \beqa
\lan:e^{-\frac{1}{2}\phi(x_1)}:e^{-\frac{1}{2}\phi(x_2)}:e^{-\phi(x_3)}\ran&=&x_{12}^{-\frac{1}{4}}x_{13}^{-\frac{1}{2}}
 x_{23}^{-\frac{1}{2}}\nonumber\\
\lan:S_A(x_1):S_B(x_2):\Psi^{\mu}(x_3):\ran&=&\frac{(\ga^{\mu})_{AB}}{\sqrt{2}}x_{12}^{-3/4}x_{13}^{-1/2}x_{23}^{-1/2}\nonumber
\eeqa
The integrand is then proportional to $x_{12}^{-1}x_{13}^{-1}x_{23}^{-1}$. For 132 ordering the integrand should be the same which fixes the following correlators:
\beqa
\lan:e^{-\frac{1}{2}\phi(x_1)}:e^{-\phi(x_3)}:e^{-\frac{1}{2}\phi(x_2)}\ran&=&x_{12}^{-\frac{1}{4}}x_{13}^{-\frac{1}{2}}
 x_{23}^{-\frac{1}{2}}\nonumber\\
\lan:S_A(x_1):\Psi^{\mu}(x_3):S_B(x_2):\ran&=&\frac{(\ga^{\mu})_{AB}} {\sqrt{2}}x_{12}^{-3/4}x_{13}^{-1/2}x_{23}^{-1/2}\labell{cor2}
\eeqa
Fixing the $SL(2,R)$ symmetry as in \reef{amp7}, one finds \reef{amp2}. This amplitude  is antisymmetric under interchanging $1\leftrightarrow 2$. This minus sign cancels  the minus sign from permutation of two fermions. So the amplitude 123+132 and 213+231 are equal. That is why one considers only non-cyclic ordering of the vertex operators in the amplitude\footnote{Alternatively, one may consider the wave function $u^A$ to be anti-commuting. In this case there is no extra minus sign when permuting two fermion vertex operators in \reef{cor10}.}. 
 The  amplitude \reef{amp2} is consistent with the non-abelian kinetic term of fermion, \ie $\Tr(\bPsi\Ga^{\mu}D_{\mu}\Psi)$. Using the observation that the S-matrix elements are independent of the pictures of the vertex operators when the internal CP factors are included, we note that the amplitude \reef{amp2} is the S-matrix element of $\lan 
V_{\bPsi}^{(1/2)}(x_1)V_{\Psi}^{(-1/2)}(x_2)V_{A}^{-2}(x_3)\ran$ in which all vertex operators have identity CP matrix. So there is no  internal CP matrix for $\Tr(\bPsi\Ga^{\mu}D_{\mu}\Psi)$, as expected, since this  coupling appears in BPS D-brane case as well.

The S-matrix element of two fermions and odd number of tachyons is given by
\beqa
{\cal A}^{\bPsi,\Psi,T,T,\cdots, T} & \sim & \int dx_1\cdots dx_{2n+3}
 \Tr\lan 
V_{\bPsi}^{(-1/2)}(x_1)V_{\Psi}^{(-1/2)}(x_2)V_{T}^{(-1)}(x_3)V_T^{0}(x_4)\cdots V_T^{0}(x_{2n+3})\ran\nonumber\eeqa
The spin operator in the S-matrix element makes the correlation function of the world sheet fermion to be non-zero. However, the internal CP factor is zero for any ordering of the vertex operators, hence,
\beqa
{\cal A}^{\bPsi,\Psi,T,T,\cdots, T}&=&0\labell{zero}
\eeqa
So there is no coupling between fermion and odd number of tachyons in the world volume of non-BPS D-breane/D-brane-anti-D-brane.

\section{The $\bPsi\Psi T^2$ amplitude}

The S-matrix element of two fermions and two tachyons is given by
\beqa
{\cal A}^{\bPsi,\Psi,T,T} & \sim & \int dx_1\cdots dx_{4}
 \Tr\lan 
V_{\bPsi}^{(-1/2)}(x_1)V_{\Psi}^{(-1/2)}(x_2)V_{T}^{(-1)}(x_3)V_T^{0}(x_4))\ran\nonumber\eeqa
Performing the $X^a(x)$ correlation function using the corresponding world-sheet propagator \reef{prop} and using the correlators in the previous section, one finds that the integrand is proportional to 
\beqa
|x_{12}|^{4k_1\cdot k_2}|x_{13}|^{4k_1\cdot k_3}|x_{14}|^{4k_1\cdot k_4}|x_{23}|^{4k_2\cdot k_2}|x_{24}|^{4k_2\cdot k_4}|x_{34}|^{4k_3\cdot k_4}\times x_{12}^{-1}x_{13}^{-\frac{1}{2}}x_{14}^{-\frac{1}{2}}x_{23}^{-\frac{1}{2}}x_{24}^{-\frac{1}{2}}
\eeqa
for any  ordering. It has the $SL(2,R)$ symmetry.  Gauging away this symmetry  by fixing $x_1=0, x_3=1, x_4=\infty$ for 1234 ordering, one finds
\beqa
{\cal A}^{1234} & \sim &-\bu_1^A\ga^{\mu}_{AB}u_2^B k_{4\mu}\frac{\Gamma(-2t)\Gamma(-2s)}{\Gamma(-2t-2s)}\Tr(\lam_1\lam_2\lam_3\lam_4)\Tr(\sigma_3\sigma_2\sigma_1)
\eeqa
 The Mandelstam variables  satisfy the on-shell condition 
\beqa
s+t+u&=&-\frac{1}{2}\labell{cons}
\eeqa

For $1243$ ordering, we fix the $SL(2,R)$ symmetry by fixing $x_1=0,x_4=1,x_3=\infty$. The amplitude becomes
\beqa
{\cal A}^{1243} & \sim &-\bu_1^A\ga^{\mu}_{AB}u_2^B k_{4\mu}\frac{\Gamma(-2s)\Gamma(-2u)}{\Gamma(-2s-2u)}\Tr(\lam_1\lam_2\lam_4\lam_3)\Tr(\sigma_3\sigma_1\sigma_2)
\eeqa

For $1324$ ordering, we fix the $SL(2,R)$ symmetry by fixing $x_1=0,x_2=1,x_4=\infty$. The amplitude becomes
\beqa
{\cal A}^{1324} & \sim &-i\bu_1^A\ga^{\mu}_{AB}u_2^B k_{4\mu}\frac{\Gamma(-2u)\Gamma(-2t)}{\Gamma(-2u-2t)}\Tr(\lam_1\lam_3\lam_2\lam_4)\Tr(\sigma_3\sigma_2\sigma_1)
\eeqa
Similarly for the other three orderings. The final result is
\beqa
{\cal A} & \sim &\bu_1^A\ga^{\mu}_{AB}u_2^B k_{4\mu}\times\nonumber\\
&&\left(\frac{\Gamma(-2t)\Gamma(-2s)}{\Gamma(-2t-2s)}\left(\Tr(\lam_1\lam_2\lam_3\lam_4)\Tr(\sigma_3\sigma_2\sigma_1)-\Tr(\lam_1\lam_4\lam_3\lam_2)\Tr(\sigma_3\sigma_1\sigma_2)\right)\right.\nonumber\\
&&+\left. \frac{\Gamma(-2s)\Gamma(-2u)}{\Gamma(-2s-2u)}\left(\Tr(\lam_1\lam_2\lam_4\lam_3)\Tr(\sigma_3\sigma_1\sigma_2)-\Tr(\lam_1\lam_3\lam_4\lam_2)\Tr(\sigma_3\sigma_2\sigma_1)\right)\right.\nonumber
\\
&&\left.+i\frac{\Gamma(-2u)\Gamma(-2t)}{\Gamma(-2u-2t)}\left(\Tr(\lam_1\lam_3\lam_2\lam_4)\Tr(\sigma_3\sigma_2\sigma_1)+\Tr(\lam_1\lam_4\lam_2\lam_3)\Tr(\sigma_3\sigma_1\sigma_2)\right)\right)\nonumber
\eeqa
One can   easily check that for the choice  $\lan 
V_{\bPsi}^{(-1/2)}(x_1)V_{\Psi}^{(-1/2)}(x_2)V_{T}^{(0)}(x_3)V_T^{-1}(x_4))\ran$ the result is exactly the same as above. The amplitude is hermitian, it  is symmetric under interchanging $3\leftrightarrow 4$ and antisymmetric under $1\leftrightarrow 2$, as expected.  Note that if one does not include the internal CP factors in \reef{amp1}, the amplitude would not satisfy any of these symmetries. Performing the trace over the internal CP matrices, one finds
\beqa
{\cal A} & \sim &\bu_1^A\ga^{\mu}_{AB}u_2^B k_{4\mu}\left(\alpha\frac{\Gamma(-2t)\Gamma(-2u)}{\Gamma(-2t-2u)}-\beta \frac{\Gamma(-2t)\Gamma(-2s)}{\Gamma(-2t-2s)}+i\eta\frac{\Gamma(-2u)\Gamma(-2s)}{\Gamma(-2u-2s)}\right)\labell{amp1}
\eeqa
where $\alpha,\, \beta $ are given in  \reef{abg} and 
\beqa
\eta&=&\frac{1}{2}\left(\Tr(\lam_1\lam_3\lam_2\lam_4)-\Tr(\lam_1\lam_4\lam_2\lam_3)\right)
\eeqa
 Using the observation that the S-matrix element in different pictures should be identical when internal CP factors are included, one finds  that the above  amplitude should be the result of the S-matrix element $\lan 
V_{\bPsi}^{(1/2)}(x_1)V_{\Psi}^{(-1/2)}(x_2)V_{T}^{(-1)}(x_3)V_T^{-1}(x_4))\ran$ in which the fermions carry identity matrix and tachyons carry $\sigma_2$. So the coupling in field theory should have no CP factor, as $\Tr(\sigma_2\sigma_2)=1$.

 Now to find the field theory couplings from the above amplitude, one should first expand this amplitude. The amplitude has massless poles in all channels. However, the on-shell constraint \reef{cons} does not allow us to send all $s,t,u$ to zero. On the other hand, the massless pole in $s$- and $u$-channels would be reproduce by the $\bPsi\Psi T$ coupling which  we have shown that string theory  does not produce it.  Hence, one must  consider the following expansion:
\beqa
s\rightarrow 0,\qquad t,u\rightarrow -\frac{1}{4}
\eeqa
in terms of momenta, the expansion is
\beqa
k_1\inn k_2,k_1\inn k_3,k_2\inn k_3\rightarrow 0
\eeqa
which is a momentum expansion.

Following \cite{Garousi:2008xp}, to expand the amplitude \reef{amp1}, we rewrite it as
\beqa
{\cal A} & \!\!\!\!\sim\!\!\!\! &\bu_1^A\ga^{\mu}_{AB}u_2^B k_{4\mu}\labell{amp4}\\
&&\left(\alpha\frac{\Gamma(2u'+2t')\Gamma(\frac{1}{2}-2t')}{\Gamma(\frac{1}{2}+2u')}-\beta \frac{\Gamma(2u'+2t')\Gamma(\frac{1}{2}-2u')}{\Gamma(\frac{1}{2}+2t')}+i\eta\frac{\Gamma(\frac{1}{2}-2t')\Gamma(\frac{1}{2}-2u')}{\Gamma(1-2t'-2u')}\right)\nonumber
\eeqa
where $t'=t+1/4=-2k_2\inn k_3$ and $u'=u+1/4=-2k_1\inn k_3$. In terms of these new Mandelstam variables, the constraint \reef{cons} becomes
\beqa
s+t'+u'=0
\eeqa
Expanding the gamma functions, one finds
\beqa
{\cal A}^{\bPsi,\Psi,T,T} & = &c\bu_1^A\ga^{\mu}_{AB}u_2^B k_{4\mu}\left(\frac{\alpha-\beta}{-2s}+\labell{amp5}\right.\\
&&\left.+\sum_{n,m=0}^{\infty}\left[a_{n,m}(\alpha t'^nu'^m-\beta u'^nt'^m)+i\eta b_{n,m}(t'^nu'^m+ u'^nt'^m)\right]\right)\nonumber
\eeqa
The coefficient $b_{n,m}$ is symmetric. Some of the coefficients $a_{n,m}$ and $b_{n,m}$ are
\beqa
&&a_{0,0}=2\ln(2),\,a_{1,0}=\frac{2\pi^2}{3}+4\ln(2)^2,\,a_{0,1}=-\frac{\pi^2}{3}+4\ln(2)^2,\\
&&a_{2,0}=8\z(3)+\frac{16}{3}\ln(2)^3+\frac{8\pi^2}{3}\ln(2),\,a_{0,2}=8\z(3)+\frac{16}{3}\ln(2)^3-\frac{4\pi^2}{3}\ln(2),\nonumber\\
&&a_{1,1}=-12\z(3)+\frac{32}{3}\ln(2)^3+\frac{4\pi^2}{3}\ln(2),\,\cdots\nonumber\\
&&b_{0,0}=\frac{\pi}{2},\,b_{1,0}=4\pi\ln(2),\,b_{2,0}=\frac{2\pi}{3}(\pi^2+12\ln(2)^2),\,b_{1,1}=\frac{2\pi}{6}(-\pi^2+24\ln(2)^2),\nonumber\\
&&b_{3,0}=\frac{8\pi}{3}(\pi^2\ln(2)+4\ln(2)^3+6\z(3)),\,b_{2,1}=\frac{8\pi}{3}(12\ln(2)^3-3\z(3)),\cdots\nonumber
\eeqa
The constant  $c$ in \reef{amp5} is a normalization constant which can be fixed by comparing the massless pole with the Feynman amplitude in effective action. 
The massless pole should be reproduced by the following non-abelian kinetic terms:
\beqa
-T_p\Tr\left(\frac{2\pi\alpha'}{2}D_aTD^aT-\frac{(2\pi\alpha')^2}{4}F_{ab}F^{ba}-\frac{2\pi\alpha'}{2}\bPsi\Ga^aD_a\Psi\right)\labell{kin}
\eeqa
The Feynman amplitude is given by
\beqa
V^{a,i}(\bPsi_1\Psi_2 A)G^{ab,ij}(A)V^{b,j}(AT_3T_4)\nonumber
\eeqa
 The vertexes and propagator are
 \beqa
 V^{a,i}(AT_3T_4)&=&T_p(2\pi i\alpha')(k_3^a-k_4^a)\left(\Tr(\lam_4\lam_3\lam^i)-\Tr(\lam_3\lam_4\lam^i)\right)\nonumber\\
 V^{a,i}(\bPsi_1\Psi_2 A)&=&T_p(2\pi\alpha')\bu_1^A\ga^a_{AB}u_2^B\left(\Tr(\lam_1\lam_2\lam^i)-\Tr(\lam_2\lam_1\lam^i)\right)\nonumber\\
 G^{ab,ij}(A)&=&\frac{i\delta^{ab}\delta^{ij}}{(2\pi\alpha')^2T_ps}\nonumber
 \eeqa
 This fixes $c=8T_p$.
 
The contact terms with coefficients $a_{0,0}$ and $b_{0,0}$ are reproduced by the following couplings:
\beqa
\alpha'T_p\Tr\left(\frac{}{}a_{0,0}\left(\bPsi\ga^a\Psi TD_a T-\bPsi\ga^a\Psi D_aT T\right)+ib_{0,0}\left(\bPsi\ga^a T\Psi D_a T-\bPsi\ga^aD_aT\Psi T\right)\right)\labell{nonabcoup}
\eeqa
It is hermitian and for abelian case it is zero. Note that the above couplings are  the same order as the kinetic terms \reef{kin}. One may try to extend this amplitude to order $(\alpha')^{1+n+m}$ along the line of \cite{Garousi:2008xp} in general non-abelian case. 
We are here interested only in the abelian couplings. The amplitude \reef{amp5} becomes
\beqa
{\cal A}^{\bPsi,\Psi,T,T} & = &c\bu_1^A\ga^{\mu}_{AB}u_2^B k_{4\mu}\left((a_{1,0}-a_{0,1})(u'-s')+O(\alpha'^2)\right)\nonumber
\eeqa
The coupling corresponding to the above amplitude is
\beqa
\frac{\alpha'^2c(a_{1,0}-a_{0,1})}{4}\bPsi\ga^a\prt^b\Psi\prt_aT\prt_bT&=&2\pi^2\alpha'^2T_p\bPsi\ga^a\prt^b\Psi\prt_aT\prt_bT\labell{ttff}
\eeqa
Now consider the  action \reef{super} in which the fields are normalized to have  the same normalization as the kinetic terms \reef{kin}, \ie
\beqa
L&\!\!\!=\!\!\!&-T_pV\sqrt{-\det(\eta_{ab}+2\pi\alpha'F_{ab}-2\pi\alpha'\bPsi\ga_a\prt_a\Psi+\pi^2\alpha'^2\bPsi\ga^{\mu}\prt_a\Psi\bPsi\ga_{\mu}\prt_b\Psi + 2\pi\alpha'\prt_aT\prt_bT )}\nonumber
\eeqa
 Expansing the above action, one finds ${\it exactly}$  the on-shell two-fermion-two tachyon couplings in \reef{ttff} including its coefficient. The above action is also consistent with the amplitude \reef{zero}. It would be interesting to extend the above amplitude to non-abelian case such that it produces the non-abelian couplings of the string theory S-matrix elements. Using the numbers $a_{0,0}=2\ln(2)$ and $b_{0,0}=\pi/2$, one observes that the non-abelian couplings \reef{nonabcoup} at order $\alpha'$  is not given by the symmetric trace of natural non-abelian extension of the above Lagrangian. 
 



\end{document}